\documentclass[sigconf]{acmart}
\AtBeginDocument{%
  \providecommand\BibTeX{{%
    \normalfont B\kern-0.5em{\scshape i\kern-0.25em b}\kern-0.8em\TeX}}}

\copyrightyear{2022} 
\acmYear{2022} 
\setcopyright{rightsretained} 
\acmConference[ARES 2022]{The 17th International Conference on Availability, Reliability and Security}{August 23--26, 2022}{Vienna, Austria}
\acmBooktitle{The 17th International Conference on Availability, Reliability and Security (ARES 2022), August 23--26, 2022, Vienna, Austria}\acmDOI{10.1145/3538969.3543810}
\acmISBN{978-1-4503-9670-7/22/08}

\usepackage{acronym}[nolist,nohyperlinks]
\usepackage{verbatimbox}
\usepackage{multirow}
\usepackage{url}
\usepackage{subcaption}
\begin{document}

\title{Learning State Machines to Monitor and Detect Anomalies on a Kubernetes Cluster}

\author{Clinton Cao}
\affiliation{%
  \institution{Delft University of Technology}
  \streetaddress{Mekelweg 5, 2628 CD Delft}
  \country{Netherlands}}
\email{c.s.cao@tudelft.nl}

\author{Agathe Blaise}
\affiliation{%
  \institution{Thales SIX GTS France}
  \country{France}}
\email{agathe.blaise@thalesgroup.com}

\author{Sicco Verwer}
\affiliation{%
  \institution{Delft University of Technology}
  \streetaddress{Mekelweg 5, 2628 CD Delft}
  \country{Netherlands}}
\email{s.e.verwer@tudelft.nl}

\author{Filippo Rebecchi}
\affiliation{%
  \institution{Thales SIX GTS France}
  \country{France}}
\email{filippo.rebecchi@thalesgroup.com}


\begin{abstract}
  These days more companies are shifting towards using cloud environments to provide their services to their client. While it is easy to set up a cloud environment, it is equally important to monitor the system's runtime behaviour and identify anomalous behaviours that occur during its operation. In recent years, the utilisation of \ac{rnn} and \ac{dnn} to detect anomalies that might occur during runtime has been a trending approach. However, it is unclear how to explain the decisions made by these networks and how these networks should be interpreted to understand the runtime behaviour that they model. On the contrary, state machine models provide an easier manner to interpret and understand the behaviour that they model. In this work, we propose an approach that learns state machine models to model the runtime behaviour of a cloud environment that runs multiple microservice applications. To the best of our knowledge, this is the first work that tries to apply state machine models to microservice architectures. The state machine model is used to detect the different types of attacks that we launch on the cloud environment. From our experiment results, our approach can detect the attacks very well, achieving a balanced accuracy of 99.2\% and a $F_1$ score of 0.982.
\end{abstract}



\maketitle

\acrodef{ai}[AI]{Artificial Intelligence}
\acrodef{attack}[ATT\&CK]{Adversarial Tactics, Techniques and Common Knowledge}
\acrodef{api}[API]{Application Programming Interface}
\acrodef{awsecs}[AWS ECS]{AWS Elastic Container Service}
\acrodef{cep}[CEP]{Complex Event Processing}
\acrodef{cic}[CIC]{Canadian Institute for Cybersecurity}
\acrodef{cip}[CIP]{Cloud Infrastructure Providers}
\acrodef{cncf}[CNCF]{Cloud Native Computing Foundation}
\acrodef{cnf}[CNF]{Cloud-native Network Function}
\acrodef{cis}[CIS]{Center for Internet Security}
\acrodef{cni}[CNI]{Container Network Interface}
\acrodef{csp}[CSP]{Communication Service Providers}
\acrodef{cve}[CVE]{Common Vulnerability Enumeration}
\acrodef{cvss}[CVSS]{Common Vulnerability Scoring System}
\acrodef{dl}[DL]{Deep Learning}
\acrodef{dnn}[DNNs]{Deep Neural Networks}
\acrodef{doh}[DoH]{DNS over HTTPS}
\acrodef{dos}[DoS]{Denial of Service}
\acrodef{ddos}[DDoS]{Distributed Denial of Service}
\acrodef{dtn}[DTN]{Digital Twin Network}
\acrodef{fsug}[FSUG]{Financial Services User Group}
\acrodef{gan}[GAN]{Generative Adversarial Network}
\acrodef{gpu}[GPU]{Graphics Processing Unit}
\acrodef{gui}[GUI]{Graphical User Interface}
\acrodef{ha}[HA]{Hardware Appliance}
\acrodef{ict}[ICT]{Information Communications Technology}
\acrodef{ids}[IDS]{Intrusion Detection System}
\acrodef{ioc}[IOCs]{Indicators of Compromise}
\acrodef{iot}[IoT]{Internet of Things}
\acrodef{ipc}[IPC]{Inter-Process Communication}
\acrodef{json}[JSON]{JavaScript Object Notation}
\acrodef{k8s}[K8s]{Kubernetes}
\acrodef{lstm}[LSTM]{Long Short-Term Memory}
\acrodef{mitm}[MITM]{Man-in-the-Middle}
\acrodef{ml}[ML]{Machine learning}
\acrodef{mlp}[MLP]{Multi-Layer Perceptron}
\acrodef{nat}[NAT]{Network Address Translator}
\acrodef{nbi}[NBI]{North-Bound Interface}
\acrodef{nfv}[NFV]{Network Function Virtualization}
\acrodef{nfvcl}[NFVCL]{NFV Convergence Layer}
\acrodef{nfvo}[NFVO]{NFV Orchestrator}
\acrodef{mano}[NFV MANO]{NFV Management and Orchestration}
\acrodef{np}[NP]{Network Policy}
\acrodef{nps}[NPs]{Network Policies}
\acrodef{nvd}[NVD]{National Vulnerability Database}
\acrodef{os}[OS]{Operating System}
\acrodef{oss}[OSS]{Operations Support System}
\acrodef{osm}[OSM]{Open Source MANO}
\acrodef{pdfa}[PDFA]{Probabilistic Deterministic Finite Automaton}
\acrodef{pid}[PID]{Process ID}
\acrodef{psp}[PSP]{Pod Security Policy}
\acrodef{pta}[PTA]{Prefix Tree Acceptor}
\acrodef{qoe}[QoE]{Quality of Experience}
\acrodef{qos}[QoS]{Quality of Service}
\acrodef{rbac}[RBAC]{Role-Based Access Control}
\acrodef{rce}[RCE]{Remote Code Execution}
\acrodef{rnn}[RNNs]{Recurrent Neural Networks}
\acrodef{sdn}[SDN]{Software Defined Network}
\acrodef{sdr}[SDR]{Software Defined Radio}
\acrodef{sgd}[SGD]{Stochastic Gradient Descent}
\acrodef{tsc}[TSC]{Training Scenario Creator}
\acrodef{tss}[TSS]{Training Scenario Supervisor}
\acrodef{ue}[UE]{User Equipment}
\acrodef{vao}[VAO]{Vertical Application Orchestrator}
\acrodef{vim}[VIM]{Virtualized Infrastructure Manager}
\acrodef{vm}[VMs]{Virtual Machines}
\acrodef{vnf}[VNF]{Virtual Network Function}
\acrodef{vsoc}[vSOC]{virtual Security Operation Center}
\acrodef{yaml}[YAML]{YAML Ain't Markup Language}

\section{Introduction}
These days more devices are being connected to the Internet.
The Internet enables users to access data and services with high availability. There are a plethora of cloud services that allow users to store their data online on a cloud server. Dropbox\footnote{https://www.dropbox.com/?landing=dbv2} and Google Drive\footnote{https://www.google.com/drive/} are two well-known examples of such cloud services. More companies are shifting towards the usage of a cloud environment to provide their services to their customers. Setting up such an environment these days is also not difficult as there exist tools that can be used to quickly set up a cloud environment. A set of detailed instructions is provided by the tools to guide developers during the setup. \ac{k8s}\footnote{https://kubernetes.io/} is an example of such a tool that provides instruction on how to set up a cluster and use it to deploy different services. 

Though the tools make it easier for a company to deploy a cloud environment and use it to provide their services to their customer, it is equally important to keep this environment secure from any malicious actor that might want to gather private (confidential) data from the customers or take down the whole environment.
In this respect, microservice-based applications must be monitored at runtime to identify anomalous behaviours that occur during their operations. This requires having a constantly updated view of the actions and events happening and then comparing them to an expected model of operation. Logs and network traffic information such as NetFlow can be collected and monitored by the services running within the environment. The challenge here is to show how to deal with the large volume of log/NetFlow data that are generated by the services.

In recent years, the utilisation of \ac{rnn} or \ac{dnn} to detect anomalies that occur during the runtime of a software system has been a trending approach investigated by many researchers~\cite{Tang2016, Du2017, Naseer2018, Meng2019}. A neural network can be trained with the log data or NetFlow data produced by the applications to model the behaviour of the applications. While it has been shown that using this approach can detect anomalies with high accuracy, it is not clear how to explain the classification decisions made by the neural network. Furthermore, it is also not clear how the neural network should be interpreted (and visualized) to understand the behaviours that are modelled. Not being able to understand or explain the decisions that are made by a model could lead a company to not adopt these models in practice as there might be legal/regulatory compliance requirements that they need to follow. 

On the contrary, is it easier to understand the behaviour that is modelled using a state machine similar to what is done for the design and analysis of software~\cite{Lee1996}. A state machine is a mathematical model used to model sequential behaviour and can be learned automatically from the traces produced by any software system. Based on a given input sequence, one can understand in which state a complex system will end up. Hence, state machines provide insights into the inner working of a software system and the models can be used to verify whether the software system is behaving as it should be~\cite{Lee1996, Clarke1999}. Extending such a state machine approach to microservice architectures is currently an open challenge.

When it comes to the learning of state machines, there are two approaches: the active-learning approach and the passive-learning approach. In the former, the learning algorithm continuously queries the system to generate training data for the learning process. LearnLib~\cite{learnlib} is a well-known framework that utilizes active learning to learn state machines. In the latter, the algorithm does not query the system to generate training data, but it simply observes the data that is produced by the system. The observations are then used for the learning of a state machine.

To detect the unusual behaviour using state machines, a probabilistic state machine can be employed. The probabilistic state machine can compute the probabilities for the traces that were produced by the system. Traces with low probabilities are considered to be unusual and an alarm should be raised. One effective method for learning probabilistic state machines in the passive-learning approach is the state-merging heuristic method.

In this work, we provide a novel anomaly detection approach that uses a passive-learning approach to learn probabilistic state machines from the NetFlow data gathered from a \ac{k8s} cluster. The state machines are used to model the normal network behaviour of the services running with the cluster. Large deviations that are detected within the models are considered to be suspicious and an alarm should be raised. We show that we can use state machines to model the runtime behaviour of a \ac{k8s} cluster and detect attacks that were launched against the cluster.

The remainder of this paper is structured as follows.
Section~\ref{section:related_work} addresses related work and positions our work.
Section~\ref{section:background} introduces different approaches to model attacks in \ac{k8s}.
Section~\ref{section:methodology} details the approach using the learning of probabilistic state machine models to detect anomalies in the cluster.
Section~\ref{section:cluster_setup} provides details on the microservice orchestration testbed that we designed, the multi-step attack scenarios deployed, and the creation of a labelled dataset.
Section~\ref{section:experiments_results} presents the evaluation of the methodology, including the detection results that we obtained.
Finally, Section~\ref{section:conclusion} draws our conclusions and points out future work.

\section{Related Work}\label{section:related_work}
In this section, we focus our attention on the relevant literature in the field of \ac{k8s} security management and the general approaches, including state-merging algorithms.

\subsection{Kubernetes security management}
While simple applications may be composed of only a few tens of components, the largest ones, such as those operated by Netflix and Uber, can run hundreds or thousands of containers. Their composition leads to complex interaction patterns that can be a source of security issues. This complexity is well understood by professionals, who require a thorough security assessment of both \ac{k8s} clusters and microservice behaviours.

By limiting the scope of our analysis to \ac{k8s}, different approaches exist to monitor the run phase of a microservice lifecycle, with the objective to detect anomalous behaviours at runtime. Application security in runtime is very different from the build and deployment stages. Dynamic analysis is in charge of assessing the security level in a \ac{k8s} cluster running in production. Compared to the analysis of monolithic applications, a microservice analysis must monitor a large number of services, each one of them running on different clusters possibly hosted on heterogeneous application platforms. Tools in this category also allow the user to define security policies, and prevent specific vulnerabilities to be exploited, providing networking segmentation and access control. Some techniques only rely on a set of signatures, while hybrid techniques use both signatures and \ac{ml} applied to logs, traffic, or both.   

As a summary, Table~\ref{tab:tools} lists the main tools for performing dynamic analysis in a \ac{k8s} cluster.
Falco~\cite{Falco} detects unexpected application behaviour and alerts on threats at runtime as defined by a customizable set of rules.
Aqua Container Security Platform~\cite{AquaSecPlatform} monitors orchestration at runtime and includes a configurable firewall.
kAudit~\cite{kaudit} and kArt perform log analysis to spot unusual or suspicious network behaviour. A firewall is embedded to allow the collection of traffic for policy compliance and anomalous behaviours detection based on \ac{ml}.
Cilium~\cite{cilium} observes network security, to enforce network policies, networking segmentation, load balancing or security policies at the kernel level.
Calico cloud~\cite{calico} provides intrusion and malware detection.
Qualys Cloud Platform~\cite{qualys} provides visibility over running services through the collection of communication data. It detects anomalous container behaviours, file access, or process activity.

Such tools have a number of shortcomings that reduce their effectiveness. First, the effectiveness of signature-based tools depends on the completeness of their definition and they may not detect zero-day attacks whose signature is not already known. Also, the commercial tools using \ac{ml} are not open-source and the methods they use are quite opaque.

\begin{table}
\centering
    \begin{tabular}{|p{2.25cm}|p{1.3cm}|p{0.9cm}|p{1.3cm}|p{.70cm}|} 
    \hline
    \textbf{Tool}                         & \textbf{Company} & \textbf{Open-source} & \textbf{Type}         & \textbf{Use eBPF}  \\ 
    \hline
    \textit{Falco}~\cite{Falco}                       & Sysdig           & Yes                  & Signatures            & Yes                \\
    \hline
    \textit{Container Security Platform}~\cite{AquaSecPlatform} & Aqua             & No                   & Signatures and ML     & No                 \\
        \hline
    \textit{kAudit kArt}~\cite{kaudit}                 & Rapid7 (Alcide)  & No                   & Signatures and ML     & No                 \\
    \hline
    \textit{Cilium}~\cite{cilium}                      & Isovalent        & Yes                  & L7 policy enforcement & Yes                \\
    \hline
    \textit{Calico cloud}~\cite{calico}                & Tigera           & Yes                  & Signatures and ML     & Yes                \\
    \hline
    \textit{Qualys Cloud Platform}~\cite{qualys}       & Qualys           & No                   & Signatures and ML     & No                 \\
    \hline
    \end{tabular}
    \caption{Classification of state-of-the art tools for dynamic analysis.}
    \label{tab:tools}
\end{table}

\subsection{General approaches}
Several approaches exist in the literature to detect malicious behaviours in microservice architectures, including signature-based approaches, \ac{ml} approaches, and state-merging algorithms.

\subsubsection{Signature-based approaches} Dynamic analysis in the microservice architecture can be performed by monitoring systems, similar to intrusion detection systems (IDS) that monitor all traffic exchanged by microservices for signs of malicious activity or security policy violations. The signature-based analysis is performed for identifying known threats, employing a set of signatures for known threats, and their connected \ac{ioc}. However, these tools are not specific to the microservice world, and usually, they do not consider the specifics of a \ac{k8s} deployment, e.g., network policies, and pod security policies, which can be beneficial to provide a context to the analysis.

\subsubsection{\ac{ml} approaches} On the other hand, ML-based approaches provide anomaly detection either via neural network or state machines approaches that can be effectively applied to learn base characteristics and identify risky deviations in behaviour. In recent years, the utilisation of neural networks has been used to detect anomalies occurring during runtime. However, it is not clear how to explain the classification decisions made by the neural network.

\subsubsection{State-merging algorithms}
State-merging algorithms first construct a \ac{pta} and then iteratively go over the states of the \ac{pta} to check which states can be merged to form a smaller state machine model. Different heuristics can be used to compute the similarities between states and to decide which similar states should be merged. There exist several tools that use state-merging algorithms to learn state machine models. 

CSight is a tool that uses state-merging algorithms to a learn state machine model from partial log data gathered from concurrent systems. The state machine model is used to model the behaviour of a distributed system~\cite{beschastnikh_2014}.

Flexfringe is another tool that uses state-merging algorithms to learn state machine models from input data. It provides a list of heuristics that can be used to decide which states should be merged. The tool allows the user to use their own heuristic so that they can compare the performance of their own heuristic to the one that is already provided~\cite{flexfringe}.

MINT uses state-merging algorithms to learn extended state machine models from input data. The difference between an extended state machine model and the definition that we provided in this paper is that a transition would only occur on an extended state machine model if several conditions are satisfied instead of one single condition. MINT also uses Genetic Programming to reduce the number of transition computations between the states during the learning process~\cite{walkinshaw_2016}.

Several works proposed different state-merging algorithms; Recently, Mediouni et al. provided an improvement on the RTI algorithm that was proposed in~\cite{verwer_rti} and with this improvement, the algorithm can learn more accurate models~\cite{mediouni_2017}. Pastore et al. proposed the TkT algorithm that computes $k$ number of transitions for each state $s$ from the initial state machine model. The $k$ transitions are considered a future transition for a state $s$ and if two states have the same $k$ transitions, these states are then merged together~\cite{pastore_ktail}. Vodenčarević et al. proposed an algorithm that learns a hybrid state machine model by taking different aspects of a system into account during the learning process~\cite{vodencarevic_2011}.

\subsection{Our contribution}
With respect to related works, our methodology can be assimilated to using state-merging algorithms to learn behavioural models from log data gathered from the runtime of microservice applications. Our models can be used to gather insights, context and explanations on why a particular behaviour has occurred during the runtime of a system. This can be crucial for a system administrator that needs to understand whether their system is being attacked or that there is a bug in their system. Additionally, we show that we can learn a state machine model that models the runtime behaviour of a system with a microservice architecture. Furthermore, our approach uses open-source tools, which gives transparency to its inner working. The overall objective is to be able to reconstruct a baseline interaction model at runtime for the microservice applications, and then interpret deviations from the model as anomalies and potential attacks. Finally, we provide a new labelled dataset, containing both malicious and normal network data, that can be used by researchers to evaluate their methods on data that is collected from a \ac{k8s} setup. 

\section{Threat modelling in Kubernetes}\label{section:background}
Microservice architectures including \ac{k8s} are composed of hundreds of containers with complex interaction patterns that can be a source of security issues. Different models have been designed to provide a taxonomy of the threats that may target such platforms.

\subsection{Microservices architectures}
In a microservices architecture, the starting point of an attack may be achieved by exploiting a misconfiguration or a vulnerability on the orchestration platform (e.g., \ac{k8s}), or compromising a vulnerability within an application (e.g., in a container).
In \ac{k8s}, platform-related vulnerabilities include a vulnerable cluster configuration, a kubelet's unauthenticated, an attacker joining as a fake worker node, \ac{rbac} issues, vulnerable network endpoints, or vulnerable service tokens. The attacker got access to the network and can then move within the cluster. Application-related vulnerabilities include misconfigured Docker, malicious Docker images, and vulnerable software applications within a pod. An application that has been compromised then leads to a foothold into the container.

With the functioning of traditional networks in mind, intuitively, such attacks are supposed to remain confined to the initially compromised pod as network-security measures are in place. Yet, exploiting the connectivity of \ac{k8s} components and the software isolation of resources, the cluster may still be compromised.

\subsection{Kubernetes threat matrices}
These sophisticated attacks usually consist of several steps that may require the detection of different indicators (e.g., malicious network traffic, CPU overload, mounting sensitive directories) to identify specific attack patterns. As of today, the correlation of data from different sources remains complex.
The MITRE \ac{attack} (adversarial tactics, techniques, and common knowledge) framework is an up-to-date database of attack techniques grouped by objectives known as \textit{tactics}. The \ac{attack} framework, originally generic, has been specialised for \ac{k8s}, highlighting multi-stage cyber-attack patterns and techniques that attackers can use to infiltrate and perform damage on \ac{k8s} clusters~\cite{Microsoft2020}. The threat matrix is made up of ten distinct families of tactics that an attacker can combine to reconstruct and exploit possible attack paths, e.g., lateral movement and privilege escalation.

Up to our knowledge, the \ac{k8s} threat matrix has been very rarely used in the literature as a means to efficiently design systems considering real-world attack scenarios. Authors in~\cite{Minna2021} and~\cite{Massoud2021} introduce several attack scenarios derived from the attack life-cycle introduced in the \ac{k8s} threat matrix.

The \ac{fsug} from the \ac{cncf} proposed a threat modelling in a \ac{k8s} cluster with attack trees~\cite{cncf-threat}. From their analysis, different end goals from an attacker have been identified:
\begin{itemize}
    \item \textit{Establishing foothold and persistence in the cluster}. The goal is to maintain persistence in a \ac{k8s} system, by compromising the platform and ensuring an ability to persist without detection. Different levels of resilience exist, ranging from none, resilience to container restart, pod deletion, node restart, and entire cluster recycling;
    \item \textit{Causing a denial of service towards the cluster}. The attacker aims to exhaust computing resources in the cluster, disrupt networking, or blackhole application traffic;
    \item \textit{Running a malicious code in workload}. The attacker aims to inject malicious code into a container, deploy poisoned container images, or write new workloads into the etcd server;
    \item \textit{Reading sensitive data}. The goal is to read or delete sensitive data, e.g., dumping the host memory, reading in the pod cache, or reading from the file system on the \ac{k8s} host.
\end{itemize}

\section{Methodology}\label{section:methodology}
The novel state anomaly detection approach presented in this work learns a state machine model from the NetFlow data collected from the \ac{k8s} cluster. As it is an open challenge for using this approach on microservice architectures, we formulate the following research questions:

\begin{itemize}
    \item \textit{RQ1: How can state machine models be used for anomaly detection on a \ac{k8s} cluster?}
    \item \textit{RQ2: How effective are the state machine models for detecting anomalies on a \ac{k8s} cluster?}
\end{itemize}

We first answer \textit{RQ1} in the following subsections.

\subsection{State-Machine Learning}
For this work, we use Flexfringe to learn \ac{pdfa}s from the NetFlow data gathered from a \ac{k8s} cluster as this framework provides us with the ability to use different state-merging heuristics for the learning of \ac{pdfa}s. The goal is to use Flexfringe to learn a \ac{pdfa} that models the normal behaviour of the \ac{k8s} cluster. Any large deviation in the behaviour is considered as an anomaly and an alarm is raised.

\subsubsection{Probabilistic State Machines}
State machines or automata are mathematical models that are used to model sequential behaviour and they can be either deterministic or non-deterministic. In the former one, the automaton can only switch to the next state if a symbol $a_i$ has been read from the input; thus an automaton cannot change state if no symbol has been read from the input. Furthermore given the current state $q_i$ of the automaton and the next input symbol $a_{i+1}$, there is a unique next state $q_{i+q}$ that the automaton will switch to. On the contrary, a non-deterministic automaton can switch to another state if no symbol has been read from the input and the automaton can switch to multiple different states when the next input symbol has been read. Determinism can be important when we are modelling sequential behaviour as deterministic models are easier to interpret and it is much more efficient to compute the transitions and probabilities on these models.

In this work, we use a specific variant of a state machine, namely the probabilistic state machine which is also known as a \ac{pdfa}. A \ac{pdfa} is a tuple $\mathcal{A} = \{\Sigma, Q, q_0, \delta, S, F\}$, where: 

\begin{itemize}
    \item $\Sigma$ is a finite alphabet
    \item $Q$ is a finite set of states
    \item $q_0 \in Q$ is a unique start state
    \item $\delta : Q \times \Sigma \to Q \cup \{0\}$ is the transition function
    \item $S : Q \times \Sigma \to \left[0,1\right]$ is the symbol probability function 
    \item $F : Q \to \left[0,1\right]$ is the final probability function, such that $F(q) + \sum_{a \in \Sigma} S(q,a) = 1$ for all $q \in Q$\\
\end{itemize}

\begin{figure}[!ht]
    \centering
    \includegraphics[scale=0.3]{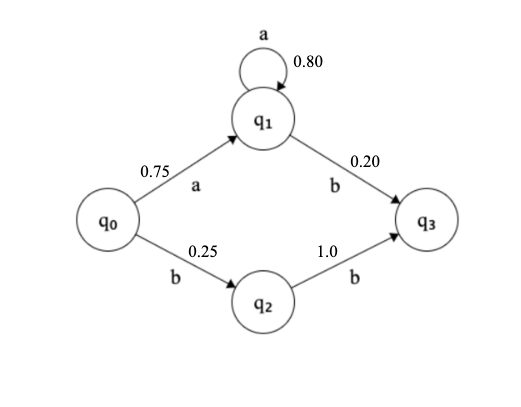}
    \caption{An example of a probabilistic state machine. Notice that each transition has a probability for it to occur and the sum of all probabilities of all outgoing transitions for a given state $q_i$ is equal to one.}
    \label{fig:example_prob_state_machine}
\end{figure}

Given a sequence of symbols $s = a_1, \ldots, a_n$, a \ac{pdfa} can be used to compute the probability for the given sequence: $\mathcal{A}(s) = S(q_0,a_1) \cdot S(q_1, a_2) \ldots S(q_{n-1}, a_n) F(q_n)$, where $\delta(q_i, a_{i+1}) = q_{i+1}$ for all $0 \leq i \leq n-1$. As an example, take the \ac{pdfa} that is shown in Figure~\ref{fig:example_prob_state_machine}. The probability for the sequence $\mathcal{A}(aaab) =  0.75 \cdot 0.80 \cdot 0.80 \cdot 0.20 = 0.096$. A \ac{pdfa} is called probabilistic because it assigns probabilities based on the symbols $S$ and final $F$ probability functions. It is called deterministic because the transition function $\delta$ (and hence its structure) is deterministic. The transition function is extended with a null symbol, representing that the transition does not exist, thus a \ac{pdfa} model can be incomplete. A \ac{pdfa} computes a probability distribution over $\Sigma^*$, i.e., $\sum_{s \in \Sigma^*} \mathcal{A}(s) = 1$. A \ac{pdfa} can also be defined without final probability function $F$, in that case it computes probability functions over $\Sigma^n$, i.e., $\sum_{s \in \Sigma^n} \mathcal{A}(s) = 1$ for all $1 \leq n$.

\subsubsection{Learning State Machine Models with State-merging Algorithms}
\begin{figure}[!ht]
    \centering
    \includegraphics[scale=0.20]{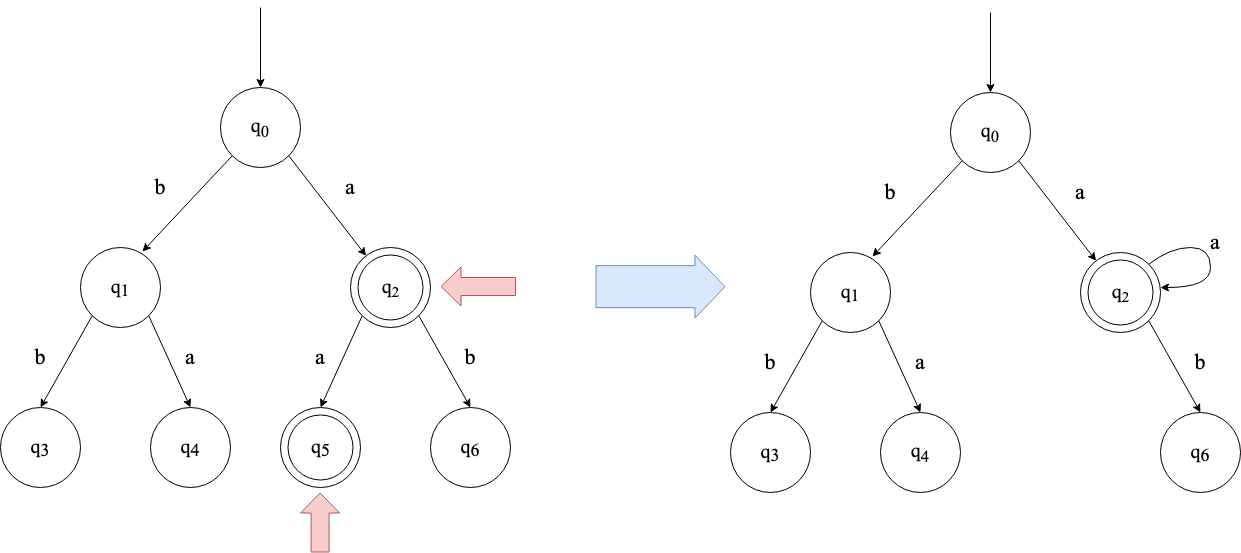}
    \caption{An example of merge performed a \ac{pta}.}
    \label{fig:example_pta_merge}
\end{figure}

One effective method for learning probabilistic state machines in the passive-learning approach is the utilisation of the state merging heuristic method~\cite{lang_1998}. This method essentially starts with a state machine that has a shape of a tree, also called a \ac{pta}. The algorithm then iteratively goes over the states of the \ac{pta} and does several comparisons between the states. A score is given for each comparison and this score is used to merge similar states, forming a smaller state machine. This process repeats until the algorithm cannot find similar states that can be merged. Figure~\ref{fig:example_pta_merge} shows an example of a merge that could occur in a \ac{pta}.

\subsection{Feature Selection \& NetFlow data Encoding (Step 1)}
For the learning of a \ac{pdfa}, we have selected three features that we used to create the traces, namely: the protocol that was used in the flow, the total number of bytes sent within the flow and the duration of the flow. We have selected these three features as we consider these three features to best describe the communication behaviour between the microservice applications and our approach does not use any service-dependent information (e.g. source/destination ports used by a service) to learn a model, making our approach more generic. Furthermore, our approach is less privacy intrusive as we try to use as few as possible features for the learning of our models.

As we are using Flexfringe as our framework for learning of \ac{pdfa}, we need to transform the raw NetFlow data into the corresponding input format of Flexfringe i.e. traces. Each flow also needs to be transformed into a single symbol. To create traces from the NetFlow data, a fixed-length sliding window can be used to create multiple traces. To transform a flow into a single symbol, it is possible to concatenate the raw values of three selected features together to form a single symbol that has the following form: $PROTOCOL\_BYTES\_DURATION$. However, using the raw values would lead to a very large \ac{pta} as the alphabet would be very large and this also increases the learning time. To reduce the learning time, we use three encoding methods: percentile encoding, frequency encoding and contextual frequency encoding. The encoding methods work as follows:

\subsubsection{Percentile Encoding}
For the percentile encoding, we take each non-categorical feature (number of bytes sent within a flow and the duration of a flow) and divide its space into bins. The encoding for a given non-categorical feature of a given flow would be the bin number that the feature value falls into. 

\subsubsection{Frequency Encoding}
For the frequency encoding, we use a threshold to decide whether a feature value occurs very frequently or not. If it is above the threshold, it is considered a frequently occurring value and we give the value a unique encoding label. If it is below the threshold, it would then be put together with other infrequent feature values and then the space is then further divided using percentiles. The drawback of using percentiles is that frequency information of the feature values is lost when the space is divided into bins. Infrequent values could be put together with frequent values and this causes the model to treat these values the same and it could cause the model to treat these infrequent values as frequently occurring values. This could lead the model to make a wrong prediction. Thus we have decided to use this encoding that takes frequency information into account when we are encoding the feature values. 

\subsubsection{Contextual Frequency Encoding}
The drawback of the frequency encoding is that it would create a very large alphabet as there could be many feature values that have a frequency that exceeds the threshold that we set. Consequently, this would create a very large \ac{pta} and thus also increases the learning time. To resolve this issue, we use an encoding that computes the counts of the next and previous feature values for each of the non-categorical feature values. The counts are then used to create a matrix and the rows are then clustered using KMeans. The clustering labels are then the encoding for the given feature. The intuition of this method is that we can use the past and future information of a flow to compute the contextual information for which a feature value occurs. Similar feature values would occur in a similar context and would therefore be clustered together, and thus also have the same encoding. With this method, we use the frequency information of the feature values and we have control over the size of the alphabet. 

\subsection{Anomaly Detection using \ac{pdfa} (Step 2)}
Once a \ac{pdfa} is learned from the NetFlow data, we can use it to compute likelihood probabilities for new traces that have been gathered during the runtime of the \ac{k8s} cluster. Whenever a trace has a very low likelihood probability, it means that the behaviour of the model deviates a lot from what is seen during training and an alarm would be raised. This trace will be then considered an anomaly. 

\section{Cluster Environment Setup}\label{section:cluster_setup}
The state-of-the-art lacks representative datasets and relevant runtime examples, mostly due to the privacy and operational specificity of deployed systems in operation. Therefore, we designed an open-source microservice orchestration testbed to collect real-world traffic and evaluate our methodology. Multi-step attack scenarios were then deployed, exploiting vulnerabilities of the platform or the applications.
We generated a labelled dataset with mixed background traffic generated by benign users and malicious traffic collected from the generation of the attack scenarios.

\subsection{Kubernetes cluster setup}
We deploy a \ac{k8s} testbed that acts as a playground on which experiments can be performed, which is easy to replicate on different platforms and adapt to different hardware requirements. The practical testbed is built on Vagrant and Ansible for reproducibility reasons, where the above scenarios can be replicated through containers and \ac{vm}. \ac{vm} are created and deployed created from a host machine in a private network, not accessible from the internet. \ac{vm} can, however, reach the Internet via a \ac{nat}.

\begin{figure}[!ht]
    \centering
    \includegraphics[width=.99\linewidth]{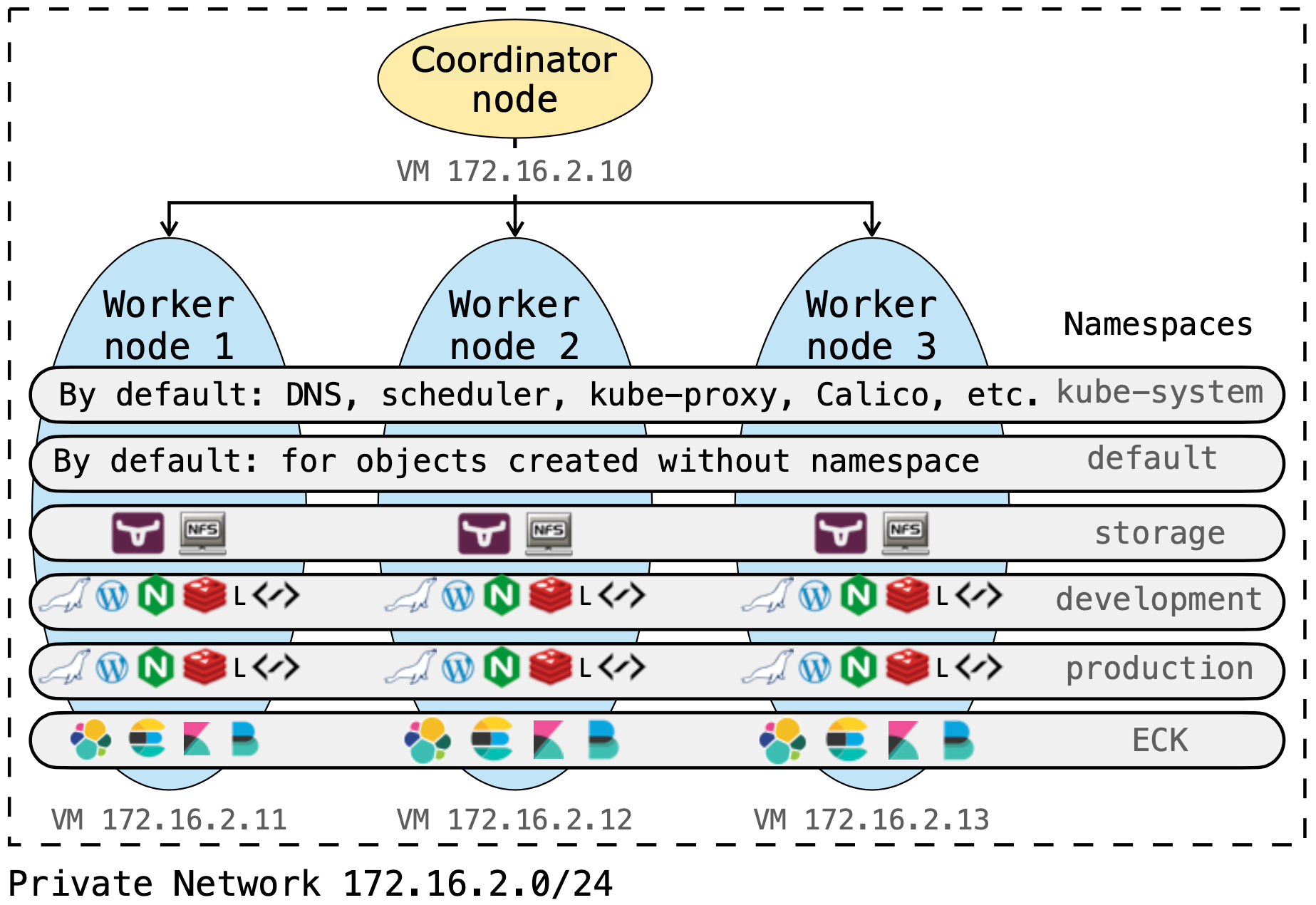}
    \caption{Application deployment on top of the Kubernetes testbed.}
    \label{fig:kubernetes-testbed}
\end{figure}

The testbed is a single-cluster setup, composed of one coordinator and at least one worker node (three in our setting), with resource isolation performed by \ac{k8s} namespaces. The applications aimed at the users are deployed on two ``development'' and ``production'' namespaces: they contain the same set of applications but with different levels of security, typically more restricted for the production namespace. Figure~\ref{fig:kubernetes-testbed} provides an overview of the multi-cluster setup and its main components.

The \textit{kube-system} namespace hosts core \ac{k8s} services, such as the DNS and the \ac{api} server. The \textit{default} namespaces host the objects created without specifying the namespace. By default, resources hosted in these two namespaces are accessible from any other namespace.
Four additional \ac{k8s} namespaces were deployed in the cluster:
\begin{itemize}
    \item \textit{Storage}: providing distributed storage through Longhorn and persistent storage through the NFS server;
    \item \textit{Development}: three applications -- a Joomla blog (with MySQL database), an OpenSSH server, and a guestbook (with a Redis leader, Redis follower, and front end);
    \item \textit{Production}: containing the same applications as the development namespace;
    \item \textit{Monitoring}: Elastic stack composed of Elasticsearch, Kibana and Beats agents (Packetbeat, Metricbeat, Auditbeat, and Filebeat).
\end{itemize}

\subsection{Description of the attack scenarios}
\begin{figure*}[!ht]
    \centering
    \includegraphics[width=.99\linewidth]{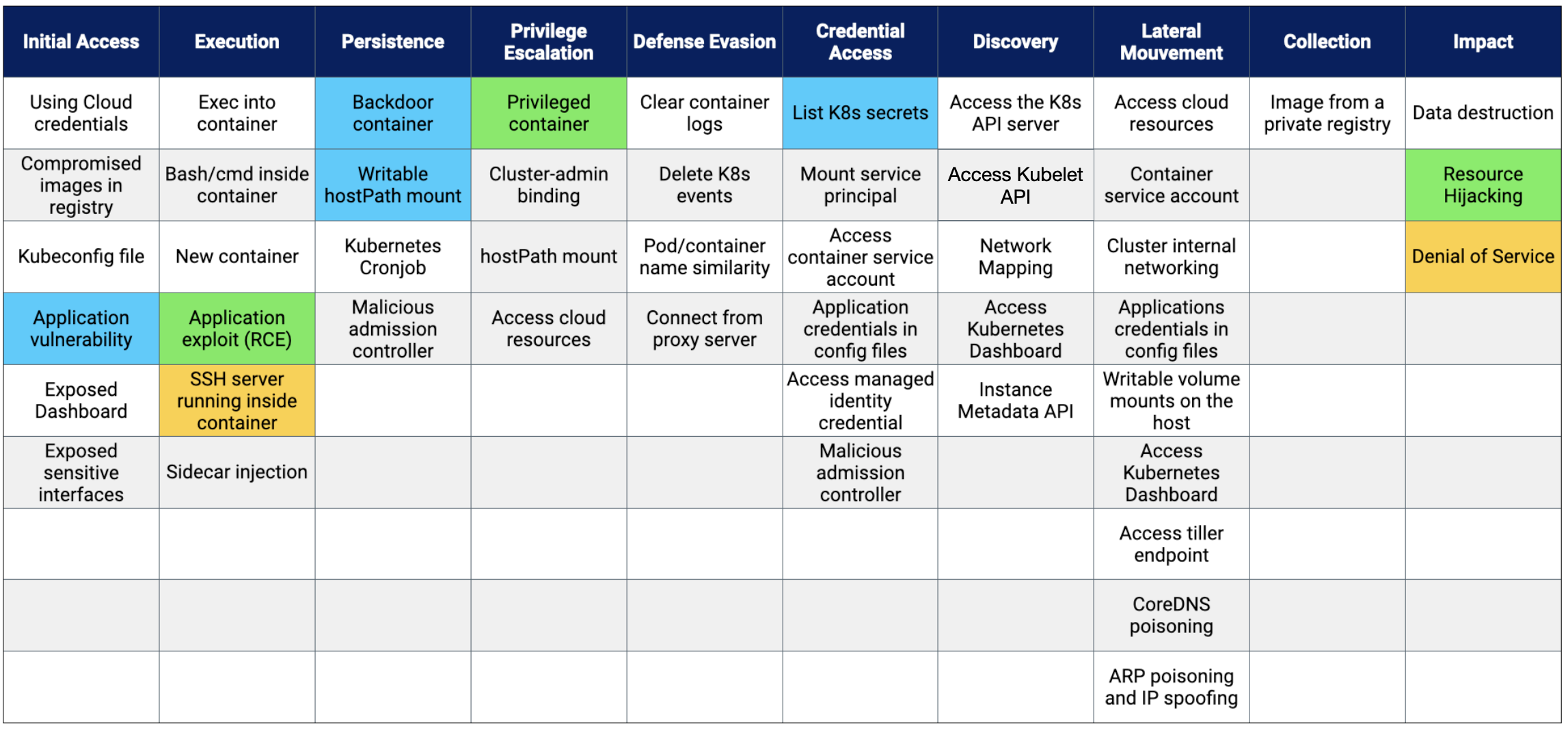}
    \caption{Threat matrix for \ac{k8s} including the three multi-step attack scenarios.}
    \label{fig:mitre_scenarios}
\end{figure*}

We propose a scenario-based view that illustrates different goals that an attacker could achieve in a \ac{k8s} cluster. Each attack scenario is fragmented into several steps to follow the \ac{k8s} threat matrix pattern. Figure \ref{fig:mitre_scenarios} introduces three sample attack scenarios related to each category in the \ac{k8s} threat matrix.

\begin{figure*}[!ht]
     \centering
     \begin{subfigure}[b]{0.2\textwidth}
         \centering
         \includegraphics[scale=0.25]{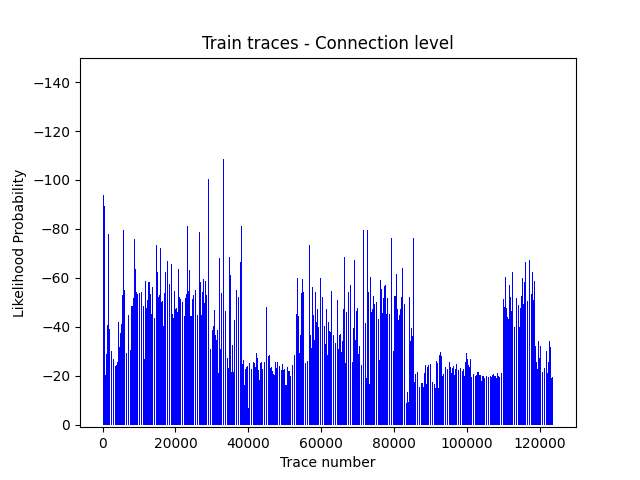}
         \caption{}
     \end{subfigure}
     \hfill
     \begin{subfigure}[b]{0.2\textwidth}
         \centering
         \includegraphics[scale=0.25]{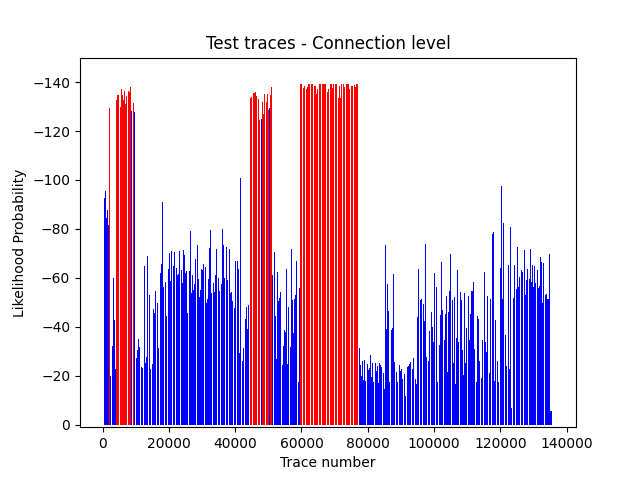}
         \caption{}
     \end{subfigure}
     \hfill
     \begin{subfigure}[b]{0.2\textwidth}
         \centering
         \includegraphics[scale=0.25]{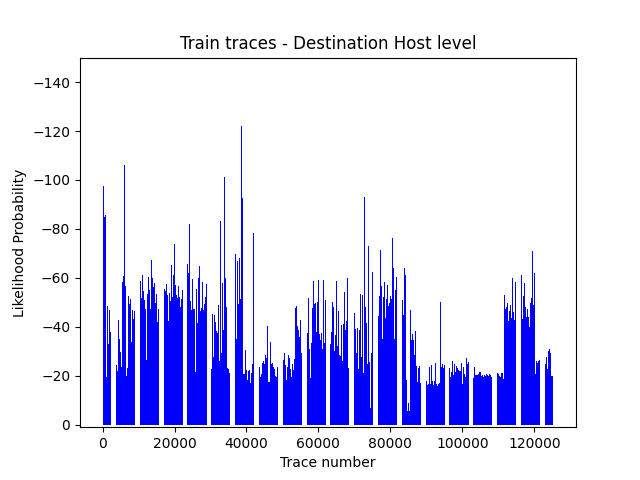}
         \caption{}
     \end{subfigure}
      \hfill
     \begin{subfigure}[b]{0.2\textwidth}
         \centering
         \includegraphics[scale=0.25]{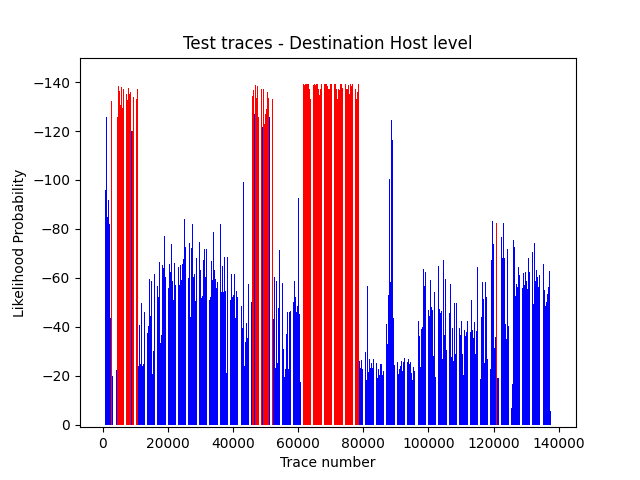}
         \caption{}
     \end{subfigure}

     \vspace{15pt}
     \hspace{-35pt}
      \hfill
     \begin{subfigure}[b]{0.2\textwidth}
         \centering
         \includegraphics[scale=0.25]{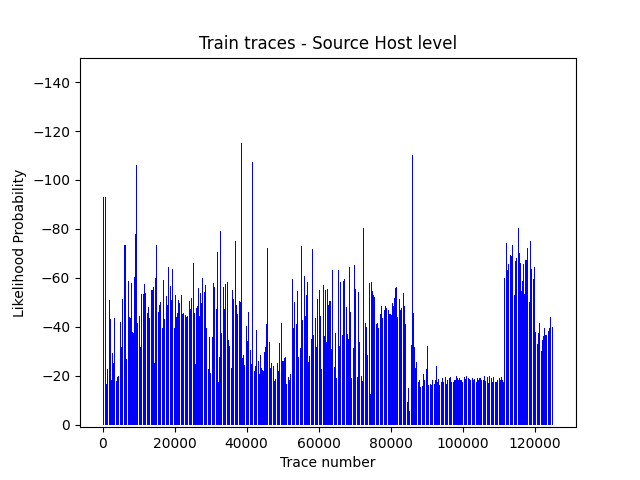}
         \caption{}
     \end{subfigure}
      \hfill
     \begin{subfigure}[b]{0.2\textwidth}
         \centering
         \includegraphics[scale=0.25]{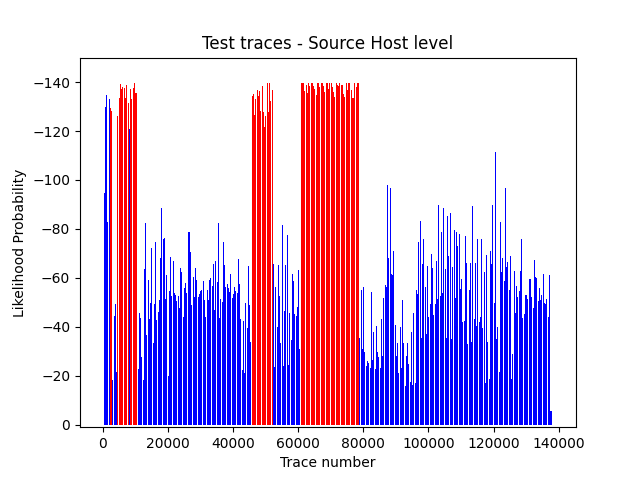}
         \caption{}
     \end{subfigure}
      \hfill
     \begin{subfigure}[b]{0.2\textwidth}
         \centering
         \includegraphics[scale=0.25]{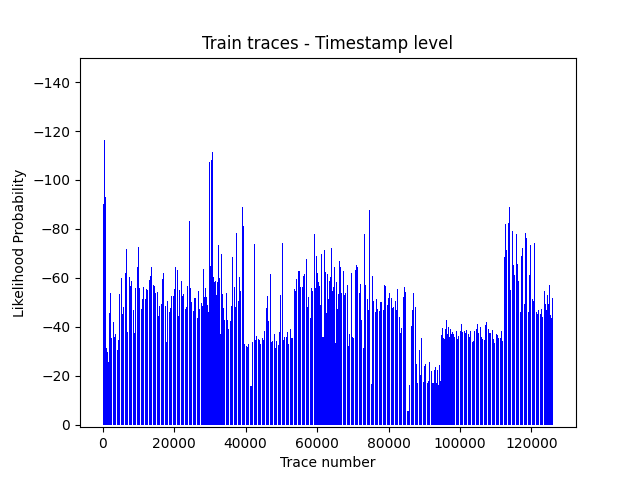}
         \caption{}
     \end{subfigure}
      \hfill
     \begin{subfigure}[b]{0.2\textwidth}
         \centering
         \includegraphics[scale=0.25]{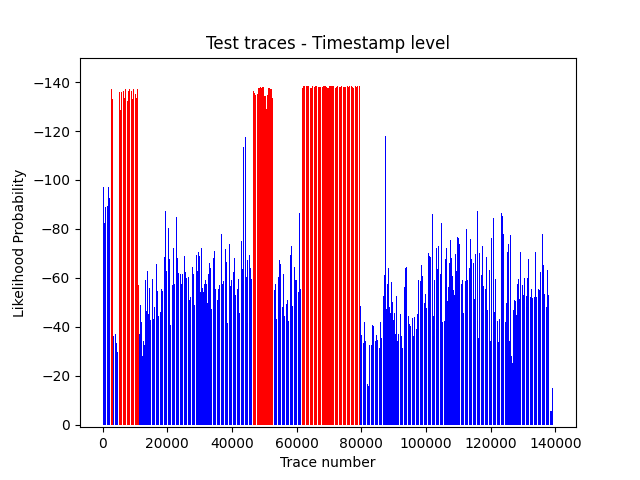}
         \caption{}
     \end{subfigure}
    
    \caption{Likelihood probability plots of the traces created using the contextual frequency encoding. The higher the peak, the lower the likelihood probability of a given trace. Subfigures $a$ and $b$ present the probabilities of the train and test traces sorted by the connections, $c$ and $d$ present the probabilities of the train and test traces sorted by destination hosts, $e$ and $f$ present the probabilities of the train and test traces sorted by source hosts, and remaining two subfigures ($g$ and $h$) present the probabilities of the train and test traces sorted by timestamp.}
    \label{fig:likelihood_plots}
\end{figure*}

\textbf{\textit{Scenario 1:} Establishing foothold and persistence.}
This scenario illustrates how an attacker can establish sophisticated pod persistence by exploiting a \ac{rce} vulnerability in a web application and getting a reverse shell. The guestbook application, composed of the front end and the Redis leader and followers, is exploited.

\begin{enumerate}
    \item \textit{Initial access}: discovering the pages of the web application and potential vulnerabilities (\texttt{dirb});
    \item \textit{Execution}: exploiting an \ac{rce} to open a reverse shell on the guestbook frontend;
    \item \textit{Discovery}: fingerprinting the type of system and environment, 
    discovering secrets mounted in the container,
    and getting the list of actions and permissions of the attacker based on its service account, e.g., get/list/exec pods located under the same namespace of the currently compromised pod.
    \item \textit{Lateral Movement}:
    the attacker uses this information to move to another pod that has root privileges.
    \item \textit{Privilege Escalation}: deploying a pod to own the node, e.g., enabling the "\texttt{hostPID}" parameter to remove the isolation and the "\texttt{nsenter}" command to give low-level access to handle the Linux kernel namespace subsystem. The attacker then breaks out of the container and gains root in the host.
    \item \textit{Persistence}:
    two variants corresponding to different persistent levels are deployed. First, a daemon set resilient to container restart but not to deployment deletion is created; then a Docker container is deployed outside the cluster, thus resilient to the entire cluster recycling.
\end{enumerate}

\textbf{\textit{Scenario 2:} \ac{dos} towards the cluster.}
This scenario illustrates a CPU \ac{dos} via the use of a vulnerable OpenSSH server. The front end of the guestbook has been built from a Docker image based on Ubuntu 16.04. This Ubuntu version includes OpenSSH v7.2, affected by \textit{CVE-2016-6515}~\cite{cve-2016-6515} that causes the CPU to overload as it does not limit password lengths for password authentication.

\begin{enumerate}
    \item \textit{Initial access}: attacker able to communicate remotely with the pod hosting the vulnerable OpenSSH server.
    \item \textit{Execution}: the remote attacker simultaneously sends 1000 requests with very long passwords to the front end pod, to cause a CPU overload.
    \item \textit{Impact}: the CPU is overloading as it receives many simultaneous requests. With the \ac{k8s} Metrics Server~\cite{metrics-server}, we identify that the CPU overload comes from the front end pod running the OpenSSH server. The memory of this pod goes from 10 to 150 MiBytes and from 5 to 550m CPU cores, only 5 seconds after launching the attack.
\end{enumerate}

\textbf{\textit{Scenario 3:} Malicious code in workload.} This scenario illustrates how an attacker can open a container shell through an \ac{rce} vulnerability in a Joomla application. The attacker then deploys malicious code in workload, e.g., a bitcoin miner in the cluster.

\begin{enumerate}
    \item \textit{Initial access}: exploiting the \textit{CVE-2015-8562} vulnerability~\cite{cve-2015-8562} on Joomla, which allows remote attackers to conduct PHP object injection attacks and execute arbitrary PHP code via the HTTP User-Agent header;
    \item \textit{Execution}: \ac{rce} attack deployed by unauthenticated remote attackers to gain instant access to the target server on which the vulnerable Joomla core version is installed.
    \item \textit{Discovery}: fingerprinting the type of system and environment and listing the running pods in the cluster.
    \item \textit{Lateral Movement}: accessing a privileged pod.
    \item \textit{Execution}: the attacker can deploy a bitcoin miner in the cluster once it gained access to a root pod.

\end{enumerate}

\subsection{Data collection}
To compensate for the lack of representative datasets, we generated our own labelled dataset collected on the microservice orchestration testbed. The objective is to collect both background traffic, produced by benign users with realistic behaviours, and malicious traffic simultaneously. A data collection session has been organised for two hours: during the first hour, thirty benign users were using the applications on the \ac{k8s} testbed, generating background traffic, and during the second hour, the attack scenarios were deployed on the cluster while benign users were still using the applications. Then, the training set is composed of traces collected during the first hour, and the test set is composed of traces collected afterwards. Traces were manually labelled depending on the IP addresses of users and the timestamps of attacks. The dataset is publically available at~\cite{assuremoss_dataset}.

\section{Experiment Results}\label{section:experiments_results}
In this section, we present the performance results of our novel anomaly detection method that uses a state machine model. The performance results that are presented in this section form the answer to \textit{RQ2}. 

\subsection{Performance Results}
The state machine model was trained on the NetFlow data that were gathered before the attacks were launched against the \ac{k8s} cluster. When creating the traces, we have sorted the NetFlow data in four different manners: connection, source host, destination host and timestamp of the flows. The intuition for sorting the NetFlow data in different manners is that some behaviour might only be seen when a particular sorting level is used. We have computed the performance results of our approach using each of the different encodings that we have mentioned in Section~\ref{section:methodology}. For comparison, we have trained an isolation forest with 100 estimators on the same training data that was used to learn our state machine model. We have used the isolation forest that is provided by the scikit-learn machine-learning package\footnote{https://scikit-learn.org/stable/modules/generated/
sklearn.ensemble.IsolationForest.html}. 

Table~\ref{tab:experiment_results} presents the performance results of our approach using the percentile encoding, frequency encoding and contextual frequency encoding, and the performance results of the isolation forest that we have trained. Based on the results that are shown in the table, our approach performs much better than an isolation forest, achieving a balanced accuracy of 99.2\% and a $F_1$ score of 0.982. 

\subsection{Likelihood Probability Plots}
In our approach, we use the state machine model to compute likelihood probabilities for the network traces and use the probabilities to assess whether there is a large deviation in the behaviour that was learned during training. Figure~\ref{fig:likelihood_plots} presents the plots of the likelihood probabilities of train and test traces for each sorting level. We only present the plots that were generated using the state machine model learned with the contextual frequency encoding as this encoding method gave the best results. For the sake of simplicity, we have coloured all malicious traces in red and all benign traces in blue. From the probabilities plots, we can see that the is a large deviation in the network traces whenever an attack has been launched against the \ac{k8s} cluster. We also see that the normal network traces produce similar probabilities as what was seen during training. From these plots, we can see why our model can detect attacks very well; visually, it is very obvious that some strange behaviour is occurring in the network of the cluster.

\begin{table}[!ht]
\centering
\caption{Performance results of our approach}
\label{tab:experiment_results}
\begin{tabular}{|c|clcc|} 
\toprule
\multicolumn{1}{|l|}{Method}                                                & Encoding                                                                                   & \multicolumn{1}{c}{\begin{tabular}[c]{@{}c@{}}Sorting \\Level\end{tabular}} & \begin{tabular}[c]{@{}c@{}}Balanced \\Accuracy\end{tabular} & \begin{tabular}[c]{@{}c@{}}$F_1$ \\Score\end{tabular}  \\ 
\hline
\multirow{12}{*}{\begin{tabular}[c]{@{}c@{}}Our~\\Approach\end{tabular}}    & \multicolumn{1}{l}{\multirow{4}{*}{Percentile}}                                            & Connection                                                                  & 0.888                                                       & 0.714                                                  \\
                                                                            & \multicolumn{1}{l}{}                                                                       & Source Host                                                                 & 0.883                                                       & 0.702                                                  \\
                                                                            & \multicolumn{1}{l}{}                                                                       & Destination Host                                                            & 0.884                                                       & 0.704                                                  \\
                                                                            & \multicolumn{1}{l}{}                                                                       & Timestamp                                                                   & 0.885                                                       & 0.715                                                  \\ 
\cline{2-5}
                                                                            & \multicolumn{1}{l}{\multirow{4}{*}{Frequency}}                                             & Connection                                                                  & 0.828                                                       & 0.689                                                  \\
                                                                            & \multicolumn{1}{l}{}                                                                       & Source Host                                                                 & 0.817                                                       & 0.675                                                  \\
                                                                            & \multicolumn{1}{l}{}                                                                       & Destination Host                                                            & 0.815                                                       & 0.673                                                  \\
                                                                            & \multicolumn{1}{l}{}                                                                       & Timestamp                                                                   & 0.796                                                       & 0.655                                                  \\ 
\cline{2-5}
                                                                            & \multirow{4}{*}{\begin{tabular}[c]{@{}c@{}}Contextual \\Frequency\end{tabular}} & Connection                                                                  & 0.991                                                       & 0.975                                                  \\
                                                                            &                                                                                            & Source Host                                                                 & 0.988                                                       & 0.967                                                  \\
                                                                            &                                                                                            & Destination Host                                                            & 0.989                                                       & 0.972                                                  \\
                                                                            &                                                                                            & Timestamp                                                                   & 0.992                                                       & 0.982                                                  \\ 
\hline
\multirow{4}{*}{\begin{tabular}[c]{@{}c@{}}Isolation \\Forest\end{tabular}} & \multirow{4}{*}{N/A}                                                                       & Connection                                                                  & 0.374                                                       & 0.0006                                                  \\
                                                                            &                                                                                            & Source Host                                                                 & 0.375                                                       & 0.0006                                                  \\
                                                                            &                                                                                            & Destination Host                                                                 & 0.373                                                       & 0.0006                                                  \\
                                                                            &                                                                                            & Timestamp                                                                   & 0.373                                                       & 0.0006                                                 \\
\bottomrule
\end{tabular}
\end{table}

\section{Conclusion \& Future Work}\label{section:conclusion}
In this work, we presented a novel anomaly detection method that uses a state machine model to detect anomalies within a \ac{k8s} cluster. To the best of our knowledge, this work is the first that tries to apply state machine models to microservice architectures. The state machine model was learned from NetFlow data gathered from the \ac{k8s} cluster when it was used normally. We have set up an experiment in which we launched three different attack scenarios on the cluster. Using our state machine approach, we managed to detect all three attack scenarios, achieving a balanced accuracy of 99.2\% and a $F_1$ score of 0.982.

For our future research directions, we would like to use other types of log data gathered from the cluster to learn our models as currently we only use the NetFlow data. Different types of log data contain different types of information that are useful to learn a more accurate model. We can assess whether using another type of log data or a combination of different log data can help us learn a more accurate model. 

Furthermore, we also want to use other datasets to evaluate whether our approach is generalisable and can be applied to other types of systems as for this work we have only used data that were collected from a \ac{k8s} cluster to run our experiments. This future research direction also allows us to compare our approach to other existing approaches.

\begin{acks}
This work is funded under the Assurance and certification in secure Multi-party Open Software and Services (AssureMOSS) Project, (\url{https://assuremoss.eu/en/}), with the support of the European Commission and H2020 Program, under Grant Agreement No. 952647.
\end{acks}

\bibliographystyle{ACM-Reference-Format}
\bibliography{references}

\end{document}